# An Electric Vehicles Migration Framework for Public Institutions in Developing Countries

ER RAQABI El Mehdi[1, 2, 3], Wenkai Li[1]

1. Graduate School of International Management, International University of Japan, Niigata 949-7277, Japan
2. Operational Researcher, Fostergy Technologies Inc., Montreal, Canada
3. Department of Mathematical and Industrial Engineering, Polytechnique Montreal, H3T 1J4, Canada

**Abstract**

Electric vehicles (EVs), with smaller environmental footprint than traditional gasoline vehicles or hybrids, are growing rapidly worldwide. Several countries such as Norway and Canada have successfully established their EV networks and achieved a significant progress towards their EV deployment. While the new EV technology is becoming popular in developed countries, emerging countries are lacking behind mainly because of the huge investment hurdle to establish EV networks. This paper provides an efficient mathematical model aiming to minimize the total costs involved in establishing an EV network, using real world data from Morocco. A given set of public institutions having a fleet of EVs are first grouped into zones based on clustering algorithms. MILP (Mixed Integer Linear Programming) models are developed to optimally select charging station locations within these organizations, with an objective to minimize the total cost. This paper can help to minimize the investment needed to establish EV networks. The transition towards EV networks can first take place in cities, especially for public institutions fleets that have a fixed and known operating itinerary and schedule, followed by locations among cities. The mathematical models provided through this paper aim to enhance and foster policy makers' ability in making decisions related to the migration towards EVs.

**Chapter One: Introduction**

**1.1 Background**

The first electric vehicle (EV) was manufactured in 1832 by Robert Anderson. Still, the EVs only became practical around 1870. Throughout the remaining years, electric cars became more popular, especially among women. Before the First World War, EVs represented a third of the vehicles on streets. Subsequently, improvements on batteries were formulated by the inventor Thomas Edison. However, the discovery of cheap crude oil led to a decrease in EVs. With the volatility of oil prices worldwide, several automakers started to explore again the electrification of vehicles, but the range anxiety related to battery depletion was among the main issues. Many efforts on both the design of EVs and the charging infrastructure took place from 2000, leading to the establishment of the necessary infrastructure in several countries including Norway, Canada, and China.

EVs have been growing[1] rapidly over the last decade. China is driving the development of the EV sector followed by the USA and Norway. The number of EVs reached 2,000,000 units in 2018, and it is expected to grow significantly in the next years. Between 2017 and 2018, sales grew up by 78% in China, 34% in Europe, 79% in USA and 86% worldwide.

This indicates a prospective shift within the transportation sector worldwide as shown in Figure 1.

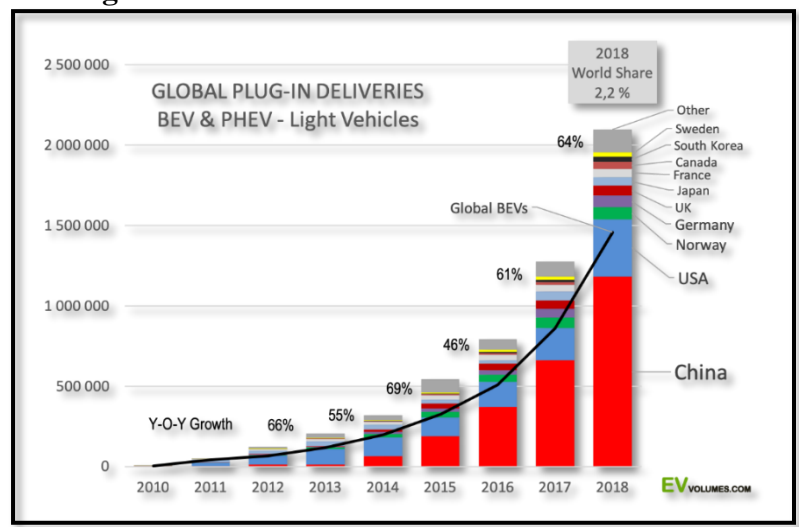

**Figure 1 - Global EV Deliveries**

As of September 2018, Norway, the country known for spreading electrification,

---

[1] Figure 1 was retrieved from www.ev-volumes.com

has achieved the highest level of EVs domination and decreased significantly emission mean. Among 10,620 new cars, 45% were entirely EVs (**Electrek, 2018**). While several countries are achieving significant advances in the incorporation of EVs, many others are still reluctant towards shifting to this environment friendly technology. Similarly to Norway, where the government is targeting electrification of all new cars within the country by 2025, and given that EV deployment has mostly been driven by policy, the migration will have a significant impact if initiated by government institutions, which are operating the public fleets. One way to tackle the risk averseness attitude is to show the significant benefits, such as cost reductions, that can be achieved through this migration.

Within this paper, centralization of EVs charging stations is presented as a prospective and efficient solution. Each government institution has a specific fleet of vehicles that is managed by a specific fitting division. Achieving the shift can be difficult, given the significant number of government institutions. In many developed countries such as Canada and Netherlands, several provinces took on greening, and especially EVs, policies and procedures for their fleets, and in some cases they are sharing experiences to support municipal governments to implement their own measures (**Greening Government Fleet, 2018**). While the deployment of EVs is advanced in these countries, it is done independently leading to a separate fleet management, which is not optimal. In fact, managing optimally EVs fleets separately is less optimal compared to managing optimally EVs fleets all at once as one fleet. Hence, the idea of centralizing the fleet management of these institutions under one organization. This organization will then manage the scheduling of EVs usage among all units. Hence, making the implementation easier in developing countries. Since these EVs will be mainly used for work purposes, their charging stations can be deployed within the workplace. Furthermore, some charging stations may also be needed on the highways. In other words, this paper will propose a framework to tackle the facility location problem:

How many charging stations are required and into which institutions they should be located? Of course, several factors take place while answering this question, namely the driving range, the cost of charging stations, and the strategic importance of the institution.

The driving range of EVs has improved significantly in recent years. Currently, the American Tesla Model S is the industry's current champion, estimated to travel a whopping 335 miles per charge with its longest-range battery pack. At the other end of the scale, the tiny two-seat Smart Fortwo Electric Drive is the EV most likely to induce range anxiety, with a paltry 59 miles per charge. As of 2018, in terms of profitability, the Chinese manufacturer BYD, the world's largest electric vehicle maker, is ahead of Tesla with a 632% increase in profits. Nowadays, the portion of EVs sold by BYD is larger than the portion of its conventional non-EV vehicles.

EVs are charged using electric vehicle supply equipment (EVSE), which are differentiated based on the level (power output range of the EVSE outlet), type (the socket and connector used for charging), and mode (the communication protocol between the vehicle and the charger). We distinguish mainly three levels of EVSEs (**Global EV Outlook, 2018**)**.** Level 1 costs less than $1000 and can add about 40 miles of range in an eight-hour overnight charge. Level 2 costs between $3500 and $6000 and can add about 180 miles of range in an eight-hour charge. Level 3 costs between $60,000 and $100,000 and can add 50 to 90 miles in half an hour. Among the three levels, level 2 chargers are the most common public chargers used within the institutions. Since our charging stations will be located in institutions, level 1 and level 2 are suitable for work charging. For highways, where time constraint is an important factor to consider, level 3 charging stations are used.

In this paper, the focus is on the design of the charging stations within a network of public institutions. Level 1 charging stations will be used for overnight charging while Level 2 charging stations will be used during working hours. Given budget limitations, charging stations cannot be opened in all the institutions. This paper's approach contains two main steps: clustering and optimization. Clusters will be constructed based on distances between institutions. Then, given the groups formed, an optimization problem is designed to identify the suitable locations for charging stations.

To identify the optimal institutions for EV charging stations, this paper first locates all the prospective locations using remote sensing software. Then, the location's output is analyzed and processed based on clustering algorithms. Finally, the clusters' output is optimized using an MIP problem. Scenario analysis will be conducted based on the problem of parameters' ranges. Once charging locations are determined, routing will be analyzed in order to ensure the efficient usage based on routing algorithms. The routing problem will be treated in further research.

This paper proposes a framework for EV network design as shown in Figure 2. First, it locates the network of institutions within a geographic area which are using remote sensing techniques. Based on the available clustering algorithms, the locations are grouped into zones. Second, locations data is extracted. Data mining techniques are necessary to prepare the input data that will be given as parameters to the mathematical model. Once completed, the clusters or zones are obtained, as well as the necessary inputs that will be used to determine optimal locations. Third, a MILP is formulated and solved using CPLEX. The three steps require several iterations to generate scenarios and comparisons. For instance, the change in the distance between institutions factor may significantly influence the number of zones and hence lead to a change within the input data. This cycle is called the learning cycle and is another advantage for decision makers who can check and implement different scenarios. The final output is a prospective migration plan towards EVs. It provides how many stations should be opened as well as their locations. In further papers, the scheduling and routing aspects will be analyzed to enhance and foster the efficient and effective usage of the available fleet of vehicles.

**Figure 2 – The Framework of EV network design**

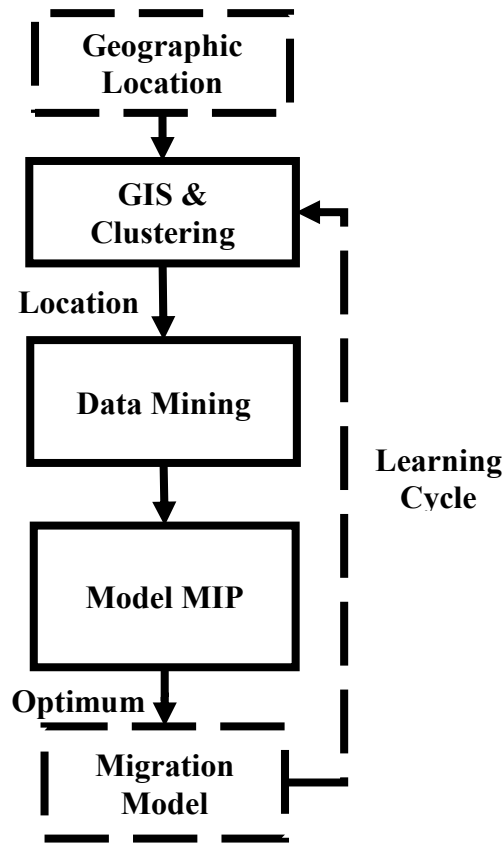

**1.2 Literature Review**

Several papers have been concerned with the facility location problem related to EV's charging stations. Most papers in the literature present models connected with level 3 charging stations, also called fast charging stations (**Hanabusa and Horiguchi, 2011; Lee et al., 2014**), which are usually required for long trips through highways. In this case, the charging demand is computed based on the number of EVs on the road as well as drivers' behavior. Some papers tackle the problem from the demand's point of view, while others tackle it based on the drivers' choices and decisions. The goal is generally the same and consists of allotting the demand to the charging stations in a balanced way. As a common tool, traffic assignment is used for the modeling of EV drivers' route choice (**Chen et al.,**

**2014**). Some studies consider gas stations' locations as a point of origin to determine the location of EVs' charging stations. However, this approach, similarly to previous ones, does not take into account the range anxiety aspect, which is among the main challenges of EVs compared to conventional vehicles.

On the other hand, some studies focused on level 2 and level 1 charging stations, also called slow charging stations (**Frade et al., 2011; Xi et al., 2013; Chen et al., 2013; J.Cavadas et al., 2015**). These stations are usually used in residential areas or in the workplace and, the models presented within these papers typically used regression analysis as a tool to estimate demand within cities. The factors considered include employment, residence, and traffic data. The goal is to therefore maximize the coverage of all EVs given the available charging stations. Regarding the coverage problem, the distinction between full coverage and partial coverage is usually missing.

On the topic of the models using a combination of fast charging and slow charging stations, few papers are available (**K.Huang et al., 2016; Z.Sun et al., 2018**). A first paper (**K.Huang et al., 2016**) considered both fast charging stations for short time needs (i.e. on highways) and slow charging stations for long time needs (i.e. within cities). It also uses the traffic assignment method. For the fast ones, the paper used a model with geometric segmentation and considers EVs moving within network links. The main parameter defined is the remaining battery capacity (driving distance). The driver should be able to find a fast charging station before complete battery depletion. For the slow ones, the demand for charging is based on zones. Within this case, the parameter defined is the walking distance. Hence, a specific point within a zone is covered if and only if the Euclidean distance between this point and the charging station is less than the maximum walking distance. The models are designed to tackle range anxiety by minimizing the total cost while guaranteeing a given level of demand coverage.

A second one also considered slow and fast charging (**Z.Sun et al., 2018**) and tackled the limited resources' constraint for both parking vehicles and vehicles on long journeys. The paper used sensitivity analysis to identify specific factors having an impact on the number and location of charging stations. These two papers found out that travelling distance as well as location's capacity are the main factors influencing the choice. Some other papers (**Liu, 2012; Liu et al., 2015**) also presented a mixed model, which first determines the number of level 1 and level 2 stations within parking and residential areas based on economic data and considers gas stations locations as a prospective location for level 3 stations.

Within the models presented above, objective functions are mainly either minimization including cost, usage, and time, or maximization including benefits and coverage.

Fast charging is crucial for solving range anxiety. However, in most emerging countries, costs incurred to acquire and manage these stations are expensive and increase the threshold of requirements of the electric grid. As a result, slow charging stations remain an efficient starting point for these countries within their cities. Then, fast charging should be added in highways to ensure the ability to manage the fleet efficiently between cities.

To design an EV charging network in emerging countries, the proposed paper highlights the importance of starting by level 1 and level 2 stations to encourage policy makers towards migration to EVs. It also includes the level 3 stations aspect when it comes to the design of the stations between cities based on the same approach.

Regarding remote sensing, it was mainly promoted for the GIS community (**Yin and Mu, 2015**). One paper (**K.Huang et al., 2016**) used auxiliary data to account for the heterogeneous distribution of demand within single polygons. The software used was ArcGIS.

Table 1 provides a summary of some of the available papers that tackled the design of electric vehicles charging stations from different perspectives. From this table, we can highlight that many papers presented the fast charging strategies and very few considered both fast and slow charging stations. The problem classification is generally an optimization problem. The goal is mainly a minimization of cost, as well as time, or a maximization of utilization, as well as coverage. The design is mainly based on a network of charging stations locations.

**Table 1 - A classification of main existing work on EV network design**

| Author(s) | Charging | Model | Goal | Design |
|---|---|---|---|---|
| **Z.Sun et al. (2018)** | Fast & Slow | Optimization | Max EV flows coverage | Nodes |
| **K.Huang et al. (2016)** | Fast & Slow | Optimization | Min total charging cost | Polygons and Links |
| **Chung and Kwon (2015)** | Fast | Optimization | Max flow captured | Graph |
| **Chen et al. (2014)** | Fast | Ad hoc | Min total travel time | Graph |
| **Lee et al. (2014)** | Fast | Optimization | Min network cost | Graph |
| **Capar et al. (2013)** | Fast | Optimization | Max flow captured | Graph |
| **Chen et al. (2013)** | Slow | Optimization | Min total access cost | Point |
| **Lam et al. (2014)** | Fast | Optimization | Min cost | Graph |
| **Liu (2012)** | Fast & Slow | Ad hoc | Min # of charging stations | Polygons and Links |
| **Hanabusa and Horiguchi (2011)** | Fast | Optimization | Min total travel time | Graph |
| **Ge et al. (2011)** | Fast | Optimization | Min flow captured | Graph |
| **Kuby and Lim (2005)** | Fast | Optimization | Max flow captured | Graph |

**1.3 Contribution**

Compared to the papers presented above, our paper is unique in four aspects:

1) It integrates remote sensing and clustering algorithms into transportation problems.

Nowadays, machine learning is extensively used in operations research. It opened a new window for better efficiency and performance in terms of execution and time reduction, which are among the most important factors in mathematical modeling. In this paper, the usage of clustering in zones definition is introduced. Based on distances between institutions, clusters of organizations are identified. Each group represents a zone and within each zone, charging stations will be opened based on several constraints. Clustering contributed significantly in reducing the execution time and making the mathematical models very powerful. While remote sensing has been used before (**K.Huang et al., 2016**), as far as we acknowledge, this article is the first to combine it with clustering. The integration of these two techniques enriches the available EVs deployment literature.

2) It tackles range anxiety problem while avoiding redundant stations.

In several previous papers, range anxiety problem is not tackled. This paper ensures that each institution belongs to one and only one zone. In that way we ensure the full coverage of all the institutions. Within each zone, a charging station is designed (either fast or slow charging). Since the fleet of electric vehicles is managed between all the institutions, the vehicle can be charged in any zone based on a specific routing and scheduling. In case a vehicle needs to be charged, another one can be used to maintain the efficiency of the network. Hence, this paper proposes a new way to tackle range anxiety issues.

3) Introduction of the importance factor approach.

As far as we acknowledge, none of the available papers used the importance factor approach. This factor is the metric used to select the optimal locations of the charging station within each zone. It is defined and computed based on the number of employees, the number of vehicles, the inflows, and the outflows. It also considers the location of the institution within the zone, i.e. the closer to the center of the zone, the higher the importance factor.

4) It is easily scalable and flexible.

Two main aspects of the model designed are scalability and flexibility. First, as shown in the paper, the model was designed within a city then scaled up to a network of three cities with two highways. The shift can be done easily and efficiently to a network of institutions within a specific country or even between countries. The same approach is used. The whole fleet is managed through all the zones based on the demand in each institution. Second, in this paper, the model for a network of public institutions is presented. Similarly, the model can be applied to any organization having a fleet of vehicles like transportation companies, travel agencies, city buses, etc. For any vehicle moving between a set of locations, this model can be applied to efficiently manage its usage and make sure it will be available and charged.

5) It enhances decision-making capability for policy makers through centralization.

For policy makers, this paper highlights the centralization aspect. Centralizing the usage of EVs between institutions has not been considered as far as we acknowledge. This approach enhances significantly the capability of decision makers since they have to deal with less design problems compared with the decentralized aspect. Using CPLEX, this paper aims to provide an optimal solution that can be adopted by policy makers to plan the migration towards EVs and the design of charging stations for a specific network of public establishments within a geographic area. It can also be used by businesses to generate insights about prospective trends in EVs. Execution time is significantly short and provides decision makers with the ability to check different scenarios and select the most suitable setting based on budget constraints, number of vehicles, and number of zones.

All these aspects considered, our paper provides a unique approach that has not been explored before in the available literature.

**Chapter Two: The Model Framework**

The model framework includes two steps. The first step is offline, called "GIS, Clustering & Data Mining". The second step is online, called "Mathematical Models".

**2.1 Step1: GIS, Clustering & Data Mining**

Before going into the model, the first and second step presented within the approach are necessary to generate inputs for the mathematical model. Since the research is conducted on geographic areas, remote sensing is crucial to manage areas efficiently. Then, once the area is located as well as the institutions within it, clustering algorithms are applied as well as some heuristics to improve solutions. Finally, the clusters obtained are added to an excel file. Several computations are conducted and tables are generated as inputs for the mathematical model.

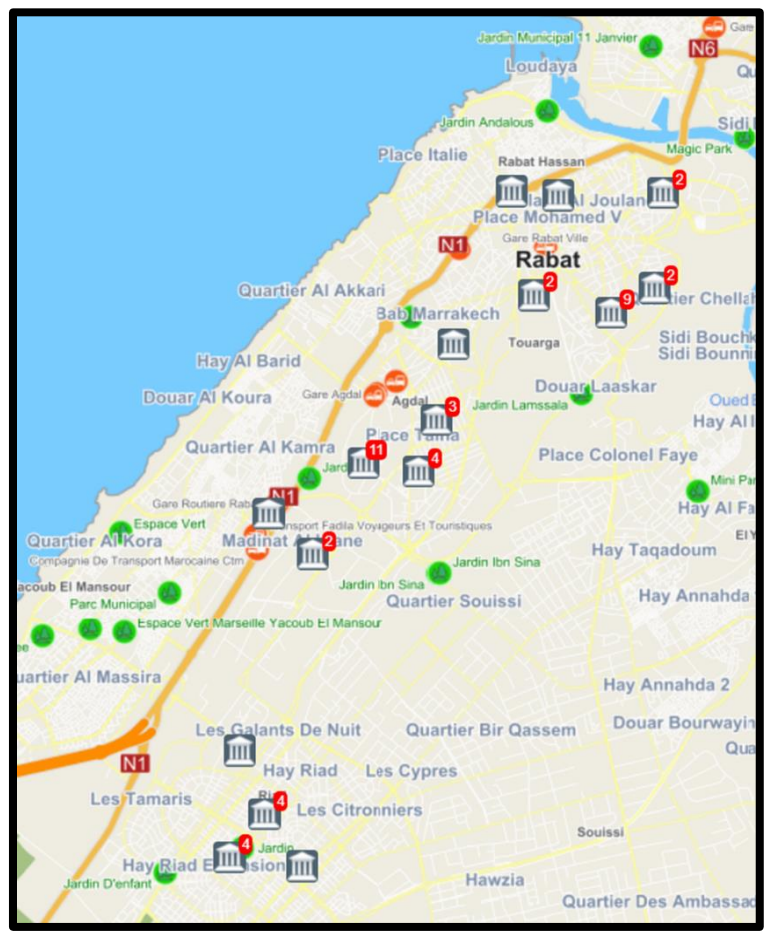

### 2.1.1 GIS

Nowadays, working with satellite pictures, Google Maps, and other geographic captures is an efficient way to conduct research in transportation. The first task is mainly identification of the geographic location, followed by the location of the set of points, or institutions in this paper's case. Usually, ArcGIS or QGIS can be used to prepare the map in case it is not available. For this paper, available maps from official Moroccan government's website were used. Figure 3 shows remote sensing for Rabat, the capital city of Morocco. The map provides the geographic area as well as the institutions' locations within the capital. The red indices in the top right corner of each blue symbol represents the number of institutions located in that area. By zooming in on the map, more details are provided and all the institutions can be identified. The information provided by the map as well as the institutions locations are the necessary inputs for the clustering step.

### 2.1.2. Clustering

In this step, connectivity-based clustering, also called hierarchical clustering, is used, which is based on the idea of grouping objects based on proximity. This is very efficient in the paper's case since it seeks to group institutions based on a normal human walking range that changes between 300m and 600m. Hence, institutions located within this range will be grouped in the same zone. Then, heuristics are used to treat some special cases such as outliers. Unique institution located around 1km and with low importance are added to the closest zone. The heuristic will simply enumerate all unique institutions, check their importance factor, and then define them either as a new zone or as an institution within the closest zone. For the previous map presented, the clustering and heuristics utilized provided the following results shown in Figure 4. The 49 institutions within the geographic area of the capital city were grouped into 7 clusters. These clusters define the zones where charging stations will be designed. It should be noted that by changing the walking range, the number of zones changes as well as the institutions' assignment.

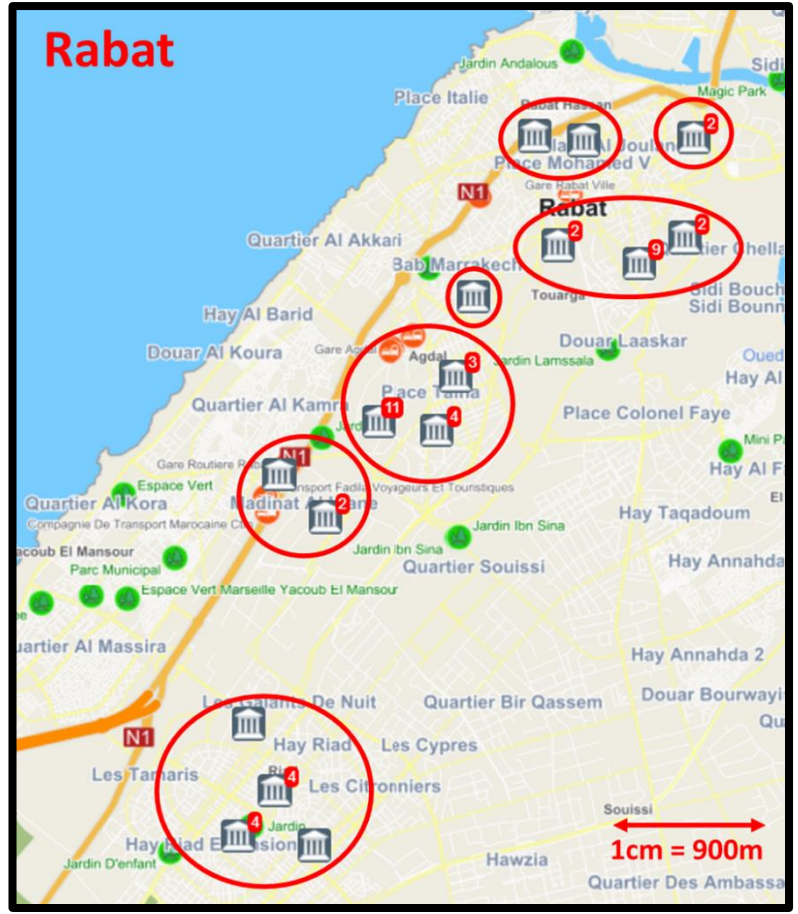

**Figure 4 - Rabat Map with Institutions Clustering**

### 2.1.3. Data Mining

Following the clustering step, the data mining step comes to prepare the necessary inputs for the mathematical model based on the previous step outputs. The Excel workbook contains seven worksheets for the MIPC model, which will be described in Step 2 below, and six for the MIPCH model, which will be described in Step 2 below, necessary to run the mathematical model. Preparing data offline is used as a technique to ensure that the programming will focus uniquely on generating the optimal solution without spending too much time on the processing and preparatory computations. This step is therefore a crucial step leading to a significant reduction of execution time necessary to identify the optimal solution.

<u>1) Worksheet 1: Institutions.</u>

This worksheet contains all the institutions considered as well as their clustering, i.e. the zone to which they belong. Each institution is given an importance factor based on the number of employees, its strategic role within the city or the highway, its location within the zone. If located in the city, the cost of level 2 charging station is computed. Otherwise, it is located on the highway and the cost of level 3 charging station is computed. These costs are also calculated based on the factors listed above.

<u>2) Worksheet 2: Clusters.</u>

This worksheet is a binary table assigning each institution to a unique zone. If the institution "i" belongs to zone "z", the excel sheet takes 1. Otherwise 0. This table supports the mathematical model in analyzing clusters.

<u>3) Worksheet 3: Importance Institutions.</u>

This worksheet simply lists the importance factor for all institutions either in the city or on the highway.

####  4) Worksheet 4: Cost Level 2 Charging Stations.

This worksheet lists the cost of level 2 charging stations if the institutions are located within the city.

####  5) Worksheet 5: Cost Level 1 Charging Stations.

This worksheet lists the cost of level 1 charging stations within each zone based on the importance factor of the zone.

####  6) Worksheet 6: Cost Level 3 Charging Stations.

This worksheet lists the cost of level 3 charging stations if the institutions are located on the highway.

####  7) Worksheet 7: Importance between zones (Only for first model).

This worksheet highlights the importance factor between zones based on the traffic between them. If the importance calculated is higher than a specific threshold, it is necessary to open a level 3 charging station.

## 2.2. Step 2. Mathematical Models

### 2.2.1. Parameters

Several parameters have been designed and computed based on the technical expertise of the involved stakeholders in the research project. These parameters were then used in the designed mathematical models. They are the following:

- ✓ **impi** = "importance factor of an institution based on location within the cluster, inflows, outflows, size, and # of employees". For each institution, we computed an importance factor. This factor depends on several variables related to each institution. The computation approach will be explained using the example of the Ministry of Economy and Finance of Morocco located in the capital city, Rabat.

**Table 2 - Ministry of Economy and Finance Data**

| Name | ID | Zones | Institution Factor | Location | Importance Factor |
|---|---|---|---|---|---|
| Ministry of Economy and Finance | i14 | z3 | 90 | M | 100 |

As shown on Table 2, the institution factor is equal to 90. This value is obtained from the four variables presented the previous section, i.e. # of employees, # of vehicles, # of inflows, # of outflows. These four variables represent the movement around the institution and consequently the usage of its fleet. For the first two variables, we score on a scale of 25. For the second two variables, we score on a scale of 20. Then we sum all the score to reach a total score on a scale of 90. Then, if the institution is located close to or within the center of the zone, we add a factor of 10 to reach the scale of 100. Hence, all institutions will have an importance factor on a scale of 100. For our example, the ministry of finance scored 25 on the first variable given the high number of employees working in it, 25 on the second variable given the high number of vehicles used to transport the employees, and 20 on both the third and fourth variables. It is also located in the middle of the zone based on our clustering algorithm. Hence, it importance factor is 100. We used the same method for all the institutions considered in our optimization model.

- ✓ **impz** = "zones importance factor": The parameter impz is the threshold between important zones and less important ones. To compute the importance factor of a zone, we take the mean of the importance factors of institutions belonging to it as follows:

$$imp_{zone} = \frac{\sum_{i \in Zone} impi_i}{|Zone|}$$

- ✓ **impbz** = "zones importance between zones": The parameter impbz is the threshold to classify zones between which the flow is quite high and others. To compute the importance between two zones, we first compute the mean of zones importances:

$$mean_{z,z'} = \frac{\frac{\sum_{i \in Zone} impi_i}{|Zone|} + \frac{\sum_{i \in Zone'} impi_i}{|Zone'|}}{2}$$

Then, we deduce the importance between the two zones based on the following formula:

$$imp_{z,z'} = \begin{cases} impbz - (100 - mean_{z,z'}) \text{ if no important flow between } z, z' \\ 100 \text{ otherwise} \end{cases}$$

Generally, within the same city, there is no need to open a level 3 charging station because zones are not very distant.

- ✓ **cl(i,z)** = "clustering of institutions i into zones z": This parameter represents the clustering of institutions into zones, the clustering is done offline, i.e. outside the model. It is binary, i.e. equals 1 if the institution I belongs to zone z and 0 otherwise.

- ✓ **c1** = "cost of building Level 1 station in zone z, this cost will be incurred if and only if the zone is very important": The parameter c1 represents the cost necessary to build a Level 1 station. Based on this definition, we can compute the cost of level 1 station for any institution as follows.

$$Cost\ Level\ 1 = \frac{\sum_{i \in Zone} impi_i}{|Zone|} \times \frac{SetupCostLvl1}{AvgImp}$$

The costs are proportional to the importance factor of the institution. In fact, the higher the importance the higher the cost since we need larger stations that can host more vehicles. Based on the field expertise, the proportional relationship is quite representative from a modeling perspective, hence the formula above. The same approach holds for the remaining costs.

- ✓ **c2** = "cost of building Level 2 station in institution i": The parameter c2 represents the cost necessary to build a Level 2 station. Based on this definition, we can compute the cost of level 2 station for any institution as follows.

$$Cost\ Level\ 2 = imp_i \times \frac{SetupCostLvl2}{AvgImp}$$

- ✓ **c3** = "cost of building level 3 station": For level 3 costs, we distinguish two cases:
  - Within the city: The parameter c3 represents the cost necessary to build a Level 3 station in the city between two zones based on the importance factor between the two zones z and z'.

  $$Cost\ Level\ 3 = imp_{z,z'} \times \frac{SetupCostLvl3}{AvgImp}$$

  - Within the highway: The parameter c3 represents the cost necessary to build a Level 3 station in the highway.

  $$Cost\ Level\ 3 = imp_i \times \frac{SetupCostLvl3}{AvgImp}$$

  Based on these definitions, we can compute the cost of level 3 station for the two cases as presented above.

### 2.2.2. MODEL 1 – MIPC

The first model is a Mixed Integer Programming (MIP) model designed for a network of institutions using a specific fleet of vehicles within a geographic area. Appendix A contains the notations used for indices, sets, parameters and variables in the mathematical formulation.

**Objective Function**

The objective is minimizing the total cost of opening a charging station within the network of institutions. The function includes the three main types, i.e. level 1, level 2, and level 3 charging stations.

$$\textbf{Minimize} \quad \sum_{i \in I} c2_i l2_i + \sum_{z \in Z} c1_z l1_z + \sum_{z,z' \in Z \cap z \neq z'} c3_{z,z'} l3_{z,z'} \quad (1)$$

Using binary variables, this paper manages where the stations should be opened. Once opened, the cost is incurred within the function and all the costs are added to compute the total cost. This paper seeks an optimal solution that could provide a benchmark to be compared with the available budget for decision makers.

**Constraints**

The model contains six constraints that ensure the selection of suitable institutions within each zone based on the importance factor presented in Appendix A. The constraints are presented below.

1) Within each zone, only one level 2 charging station can be opened:

$$\sum_{i \in I} cl_{i,z} l2_i = 1 \quad \forall z \in Z \qquad (2)$$

Based on the data input, each institution belongs to a specific zone. Hence, each zone will contain several institutions. Among all these institutions, one will be selected to be the location of the level 2 charging station, which will contain several level 2 electric terminals. All the vehicles will have to charge within the selected location inside the zone into which they operate.

2) Within each zone, the level 2 charging station will be opened in the most important institution.

$$\sum_{i \in I} cl_{i,z} impi_i l2_i \geq \max_{i \in I} cl_{i,z} impi_i \quad \forall z \in Z \qquad (3)$$

Based on the data input, the level 2 charging station will be opened in the most important institution within the zone. This is mathematically expressed with the right side of the constraint above. Since it is a minimization problem, the constraint will be satisfied with equality. The equality is ensured by the institution with the highest importance factor within the zone. The importance factor is defined based on several information including the location of the institution with the zone, i.e. the ones in the middle are given more priority, as well as the number of employees, the number of inflows and outflows, etc. Since it is a minimization problem and given the previous constraints, the model will select the equality case ensured by the most important institution.

3) Within each zone, a level 1 charging station may be opened in case the zone importance is higher than a specific threshold.

$$impz \sum_{i \in I} cl_{i,z} + (100 - impz) l1_z \sum_{i \in I} cl_{i,z} \geq \sum_{i \in I} cl_{i,z} impi_i \quad \forall z \in Z \qquad (4)$$

This paper defines also zones' importance. It is equal to the average of all the institutions' importance within the zone. A zone is considered important if its important factor is higher than a specific importance threshold. To set the threshold mathematically, we use the big M method from operations research. The smallest value of the big M is (100-impz). Within the selected zones, electric vehicles need to be charged during the whole night to be ready the day after. In this case, a level 1 charging station containing several level 1 electric terminals will be opened for night charging.

4) A level 3 charging station may be opened in case the importance between two zones is higher than a specific threshold.

$$imp_{z,z'} \leq impbz + (100 - impbz)l3_{z,z'} \qquad (5)$$

The importance between zones is defined based on the flows between zones, i.e. the # of vehicles moving between each set of two zones. In case the importance factor is higher than a specific threshold determined using the big M method similarly constraint 4, a level 3 charging station will be opened in order to ensure the continuous flow between the two zones. Within the same city, it is usually not necessary given the short distances. Still, this paper uses the constraints for special cases that may occur within a city and where it is necessary to add fast charging to avoid flow interruption.

5) Binary variables definition.

$$lk_i \in \{0, 1\} \quad \forall k \in \{1, 2, 3\} \, \forall i \in I \qquad (6)$$

Finally, the decision to open or not within a specific institution is achieved through binary variables.

**Assumptions**

The following assumptions are used for MIPC:

- EVs users can walk between the institutions located in the same zone. It means that each zone will group institutions that are within walking range of a normal human, i.e. between 300m and 600m. Consequently, the user can face three scenarios based on the schedule. First, the user may use the available EVs within his institution. Second, he may need to walk towards the charging station's institution in case there is no EV in his institution. Third, he may wait for a prospective arrival of an EV into his institution.

- EVs will be used within zones and between zones. They do not belong to any specific zone. They can be easily moved toward high demand zones and managed based on usage cycles. They can charge in any zone.
- The binary variable represents one charging station. Each charging station will contain several electric terminals of the same type selected, i.e. a charging station of level 2 will contain several electric terminals of level 2. The number of electric terminals can be determined based on the number of EVs, which is out of the scope of this paper.

**Data**

The first model is applied to a set of public institutions located in the capital of Morocco, Rabat as shown in Figure 4. The number of institutions is 49. Table 3 presents the head of the Excel worksheet. The first step of the model's framework provided the clustering of each institution in the second column. Each institution has an institution factor that is estimated based on the number of employees, inflows, outflows, etc. Then, based on the institution factor and the location of the institution with the zone, the importance factor is computed. The last column provides the estimated costs for designing a charging station within the institution.

**Table 3 - Head of the Model 1 Dataset**

| ID | Zones | Institution Factor | Location | Importance Factor | Cost Level 2 ($) |
|----|-------|--------------------|----------|-------------------|------------------|
| i1 | z1 | 70 | M | 80 | 5714 |
| i2 | z1 | 70 | M | 80 | 5714 |
| i3 | z2 | 70 | M | 80 | 5714 |
| i4 | z2 | 80 | M | 90 | 6429 |
| i5 | z3 | 90 | - | 90 | 6429 |
| i6 | z3 | 70 | - | 70 | 5000 |

**Results**

Two main factors influence the results. The first one is the walking distance. The second one is the importance threshold, but its impact is not significant enough to be considered in scenario analysis. The results are therefore computed for different sets of walking distances as presented in Table 4. The model size for the different sets is presented in Table 5.

| Walking Distance (m) | # Single Equations | # Single Variables | # Discrete Variables |
|---|---|---|---|
| 300 | 271 | 290 | 289 |
| 400 | 131 | 160 | 159 |
| 500 | 71 | 106 | 105 |
| 600 | 41 | 80 | 79 |

**Table 4 – MIPC Model Size**

| Walking Distance (m) | Sol. Time (s) | # Zones | # Level 2 | # Level 1 | # Level 3 | Total Cost ($) |
|---|---|---|---|---|---|---|
| 300 | 14 | 15 | 15 | 5 | 0 | $ 90,284.00 |
| 400 | 12 | 10 | 10 | 3 | 0 | $ 63,428.00 |
| 500 | 12 | 7 | 7 | 2 | 0 | $ 44,434.00 |
| 600 | 11 | 5 | 5 | 2 | 0 | $ 33,325.00 |

**Table 5 - MIPC Model Results**

The mathematical model was implemented using CPLEX 12 for GAMS Software and run on an HP computer with Intel Core i7 7$^{th}$ Gen.

The model results are negatively correlated with the walking distance. A higher walking distance leads to more zones and therefore more charging stations and higher costs. The execution is quite efficient with less than 15s. This is mainly due to the offline work and data mining done on Excel.

The data prepared makes computations easy and solution reaching swift, despite the high number of decision variables and constraints. Given the short execution time, the model is fostering policy maker's ability in making decisions related to the migration. By simulating different scenarios, the cost minimization provides a good benchmark to compare with the available budget for the migration.

### 2.2.1. MODEL 2 – MIPCH

The second model is also an MIP and an extension of the first one into a network of cities linked through highways. The appendix contains the notations used for additional indices, sets, parameters and variables in the mathematical formulation. The model is then applied to the capital Rabat, and two other cities Casablanca and Fes linked through two main highways within the emerging country Morocco.

**Objective Function**

The objective is to minimize the total cost of opening a charging station within the network of institutions within the cities as well as in the highways. The function includes the three main types, i.e. level 1, level 2, and level 3 charging stations.

**Minimize** $\sum_{i \in C} c2_i l2_i + \sum_{z \in Z} c1_z l1_z + \sum_{i \in H} c3_i l3_i$ \qquad (7)

Using binary variables, this paper manages where the stations should be opened. Once opened, the cost is incurred within the function and all the costs are added to compute the total cost. C denotes city while H denotes highway. Within city, we open level 2 charging stations.

In the highways, level 3 charging stations are opened. Level 1 charging stations are opened similarly to MIPC. This paper seeks an optimal solution that could provide a benchmark to be compared with the available budget for decision makers.

**Constraints**

The model contains six constraints that ensure the selection of suitable institutions within each zone based on the importance factor. The constraints are presented below.

1) Open one level 2 charging station within each zone belonging to a city.

$$\sum_{i \in C} cl_{i,z} l2_i = 1 \quad \forall z \in ZC \qquad (8)$$

2) Open one level 3 charging station with each zone belonging to a highway.

$$\sum_{i \in H} cl_{i,z} l3_i = 1 \quad \forall z \in ZH \qquad (9)$$

Based on the data input, highways are a part of the first model extension. Hence, each highway will contain several zones and each zone will contain several institutions. Among all these institutions within each zone, one will be selected to be the location of the level 3 charging station, which will contain several level 3 electric terminals. EVs moving between cities can charge rapidly during the trip before battery depletion.

3) Within each zone, the level 2 charging station will be opened in the most important institution within each zone in the city.

$$\sum_{i \in C} cl_{i,z} impi_i l2_i \geq \max_{i \in C} cl_{i,z} impi_i \quad \forall z \in ZC \qquad (10)$$

4) Within each zone, a level 3 charging station may be opened in the most important institution within each zone on the highway.

$$\sum_{i \in H} cl_{i,z} impi_i l3_i \geq \max_{i \in H} cl_{i,z} impi_i \quad \forall z \in ZH \qquad (11)$$

This paper uses the analogy with level 2 charging stations to formulate the constraints related to level 3 charging stations. The highway will be segmented into zones containing several institutions as prospective locations. Then the same selection process as level 2 will be followed.

5) Within each zone, a level 1 charging station may be opened in case the zone importance is higher than a specific threshold.

$$imp_z \sum_{i \in C} cl_{i,z} + (100 - imp_z)l1_z \sum_{i \in C} cl_{i,z} \geq \sum_{i \in I} cl_{i,z} impi_i \quad \forall z \in ZC \quad (12)$$

6) Binary variables definition.

$$lk_i \in \{0, 1\} \quad \forall k \in \{1, 2, 3\} \forall i \in I \quad (13)$$

**Assumptions**

The following assumptions are used for MIPCH:

- The highway will be segmented based on a specific distance that was fixed to 30km. It means that each 30 km, there will be a possibility to charge the EV. There will be more prospective locations towards the middle of the highway compared to the cities' areas.
- In this second model, this paper assumes that level 3 charging stations are not needed within cities and they will be implemented uniquely on the highways.
- This paper assumes that the 30km distance is sufficient enough to tackle the range anxiety with the highway. It also simulates different segmentation distances.
- EVs will be used within zones, between zones, and on highways. They do not belong to any specific city. They can be easily moved toward high demand zones and managed based on usage cycles. They can charge in any location.
- The binary variable represents one charging station. Each charging station will contain several electric terminals of the same type selected, i.e. a charging station of level 3 will contain several electric terminals of level 3. The number of electric terminals will be determined based on the number of EVs.

**Data**

The second model was applied to the capital of Morocco Rabat, Casablanca, and Fes. These three cities are linked by two highways. The number of institutions is 259 in both cities and highways. This paper applies the same approach used in Model 1 for each city and extends it for highways by adding level 3 charging stations as presented on Table 6. It is assumed that level 1 and level 2 charging stations are sufficient within cities while level 3 charging stations are required to make charging faster. The data for cities and for highways is similar to the one provided in the data section of model 1. The main difference is the level of charging stations.

**Figure 5 - Morocco Map with 3 Cities and 2 Highways**

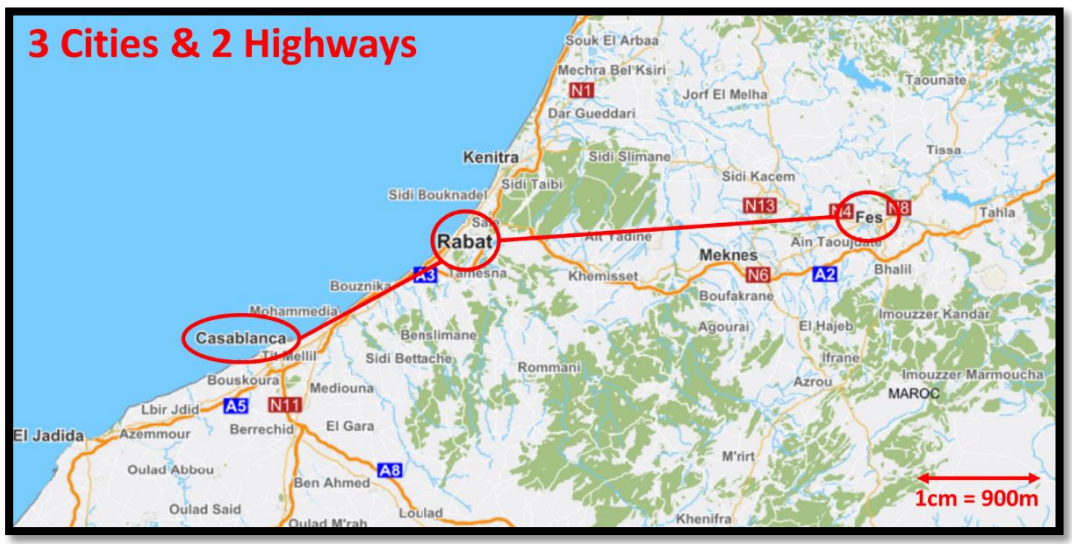

| ID | Zones | Importance Factor | Cost Level 3 ($) |
|---|---|---|---|
| i216 | z50 | 67 | 28714 |
| i217 | z50 | 81 | 34714 |
| i218 | z50 | 72 | 30857 |
| i219 | z50 | 63 | 27000 |
| i220 | z51 | 65 | 27857 |
| i221 | z51 | 65 | 27857 |
| i222 | z51 | 84 | 36000 |
| i223 | z51 | 74 | 31714 |
| i224 | z51 | 64 | 27429 |

**Table 6 - Portion of Highways Data**

### Results

Three main factors influence the results. The first one is the walking distance. The second one is the highway segmentation distance. The third one is the importance threshold, but its impact is not significant enough to be considered in scenario analysis. The results are therefore computed for different sets of walking distances and highway segmentation distances as presented on Table 7 and Table 8. The model size for the different sets is presented in Table 9 and Table 10.

Similarly, to the first model, the mathematical model was implemented using CPLEX 12 for GAMS Software and run on an HP computer with Intel Core i7 7$^{th}$ Gen.

**Table 7 - MIPCH Results for Segmentation Distance 30km**

| Walking Distance (m) | Sol. Time (s) | # Zones | # Level 2 | # Level 1 | # Level 3 | Total Cost ($) |
|---|---|---|---|---|---|---|
| 300 | 23 | 80 | 70 | 22 | 10 | $ 800,068.00 |
| 400 | 16 | 66 | 56 | 16 | 10 | $ 714,094.00 |
| 500 | 15 | 59 | 49 | 10 | 10 | $ 668,276.00 |
| 600 | 13 | 45 | 35 | 9 | 10 | $ 587,333.00 |

**Table 8 - MIPCH Model Results for Walking Distance 500m**

| Segmentation Distance (km) | Sol. Time (s) | # Zones | # Level 2 | # Level 1 | # Level 3 | Total Cost ($) |
|---|---|---|---|---|---|---|
| 15 | 19 | 70 | 49 | 10 | 21 | $ 1,020,792.00 |
| 30 | 16 | 59 | 49 | 10 | 10 | $ 656,956.00 |
| 40 | 15 | 57 | 49 | 10 | 8 | $ 598,848.00 |
| 60 | 12 | 53 | 49 | 10 | 5 | $ 457,419.60 |

**Table 9 - MIPCH Model Size for Segmentation Distance 30km**

| Walking Distance (m) | # Single Equations | # Single Variables |
|---|---|---|
| 300 | 231 | 330 |
| 400 | 189 | 316 |
| 500 | 168 | 309 |
| 600 | 126 | 295 |

**Table 10 - MIPCH Model Size for Walking Distance 500m**

| Segmentation Distance (km) | # Single Equations | # Single Variables |
|---|---|---|
| 15 | 109 | 309 |
| 30 | 168 | 309 |
| 40 | 164 | 309 |
| 60 | 156 | 309 |

Within this second model, level 3 stations are required on the highways since fast charging is necessary.

Comparably to the first model, the model results are negatively correlated with the walking distance within the city and the segmentation distance on the highway. A higher walking distance or segmentation distance leads to more zones and therefore more charging stations and higher costs. The execution is less than 25s. This is mainly due to the offline work and data mining done on Excel. The data prepared makes computations easy and solution reaching rapid despite the high number of decision variables and constraints. The second model highlights the scalability of the model since this paper was able to expand the model into a network of cities and highways starting from the single city model (MIPC) presented above.

**Chapter Three: Conclusion**

This paper suggested an efficient and effective method that can significantly help decision and policy makers in emerging countries towards the shift to EVs. For any network of public or private institutions that are using a fleet of vehicles, the solution presented, based on the centralization principle, can enhance and foster usage and management of EVs. The approach is based on an offline setup then a mathematical model. The numerical experiments of both MIPC and MIPCH to the capital of Morocco Rabat as well as the three Moroccan cities (Rabat, Casablanca, and Fes) with two highways showed the practicability of the research and the possibility of using it for benchmarking the budget necessary for the migration. Several decision makers within developing countries lack the usage of operations research in making efficient decisions. Hence, the solution provided from the optimization model can help in saving significant amount of money. This research opened another window towards an important topic that will logically follow in other papers and which is related to routing and scheduling. Since the fleet management will be centralized, adding routing and scheduling will ensure an efficient management and an effective planning of demand based on the available resources. The insight underlining this paper is about planning the routes and locations through which a specific vehicle will flow as well as its users. This can lead to a better management and utilization of charging stations as well as a good balance of demand between all zones either in the city or on the highway.

**Acknowledgments**

The authors are grateful for technical support from Fostergy Technologies Inc., which specializes in EV simulation as well as the migration from conventional vehicles towards EVs.

**Appendix**

Set and Parameters

a) Sets

**i** institutions "all the institutions within the urban network";

**z** zones "zones or groups of institutions"

b) Parameters

**AvgImp** = "cost of building level 3 station between zones": The parameter c3 represents the cost necessary to build a Level 3 station. In this paper, we assume this parameter to be equal to 70.

**c1** ="cost of building Level 1 station in zone z, this cost will be incurred if and only if the zone is very important": The parameter c1 represents the cost necessary to build a Level 1 station.

**c2** = "cost of building Level 2 station in institution i": The parameter c2 represents the cost necessary to build a Level 2 station.

**c3** = "cost of building level 3 station between zones": The parameter c3 represents the cost necessary to build a Level 3 station.

**cl(i,z)** = "clustering of institutions i into zones z": This parameter represents the clustering of institutions into zones, the clustering is done offline and is read from Excel file.

**impi** = "importance factor of an institution based on location within the cluster, inflows, outflows, size, and # of employees": The parameter impi provides the importance factor for each institution.

**impbz** = "zones importance between zones" : The parameter impbz is the threshold to classify zones between which the flow is quite high and others. In this paper, we assume this parameter to be equal to 95.

**impz** = "zones importance factor zones factor": The parameter impz is the threshold between important zones and less important ones. In this paper, we assume this parameter to be equal to 80.

**SetupCostLvl1** = "setup cost for an average level 1 station": The parameter SetupCostLvl1 represents the cost necessary to build an average level 1 station. To install such a station, the station cost is between $300 and $600 while the parts cost varies between $0 and $1700. For an average institution, i.e. an institution with an importance factor equal to 70, we assume in this paper that the level 1 station cost is 500$ ($300 for the station and $200 for the labor).

**SetupCostLvl2** = "setup cost for an average level 2 station": The parameter SetupCostLvl2 represents the cost necessary to build an average level 2 station. To install such a station, the station cost is between $500 and $2200 while the parts cost varies between $1200 and $3300. For an average institution, i.e. an institution with an importance factor equal to 70, we assume in this paper that the level 2 station cost is 5000$ ($2000 for the station and $3000 for the labor).

**SetupCostLvl3** = "setup cost for an average level 2 station": The parameter c3 represents the cost necessary to build a Level 3 station. To install such a station, the station cost is between $20000 and $50000 while the parts cost is above $10000. For an average institution, i.e. an institution with an importance factor equal to 70, the level 3 station cost estimation is 30000$ ($20000 for the station and $10000 for the labor).

c) Variables

**l1**: "binary variable =1 if level 1 station designed in zone z, =0 otherwise"

**l2**: "binary variable =1 if level 2 station designed in institution i, =0 otherwise"

**l3**: "binary variable =1 if level 3 station designed between city co and city

**tc**: total design cost ;